\documentstyle[epsfig]{mn}

\begin{document}
\title[Jeans  criterion modified by external field]
{Jeans instability criterion modified by external tidal field}

\author[C.J. Jog]
       {Chanda J. Jog\thanks{E-mail : cjjog@physics.iisc.ernet.in}\\
  Department of Physics,
Indian Institute of Science, Bangalore 560012, India \\
} 


\maketitle

\begin{abstract}
The well-known Jeans criterion describes the onset of instabilities in an infinite, homogeneous, 
self-gravitating medium supported by pressure.  Most realistic astrophysical systems, however, are 
not isolated - instead they are under the influence of an external field such as the  tidal field due to a neighbour.
Here we do a linear perturbation analysis for a system in an external field, 
and obtain a generalized dispersion relation that depends on the wavenumber, the sound speed, and also  the
magnitude of the tidal field. A typical, disruptive tidal field is shown 
to make the system more stable against perturbations, and results in a 
higher effective Jeans wavelength. The minimum mass that can become unstable is then higher (super-Jeans) than the usual Jeans mass. 
Conversely, in a compressive tidal field,
 perturbations can grow even when the mass is lower (sub-Jeans). 
This approach involving the inclusion of tidal field opens up a 
new way of looking at instabilities in gravitating systems. 
The treatment is general and the simple analytical form 
of the modified Jeans criterion obtained makes it easily accessible.
\end{abstract}
\section{Introduction}
The stability of a gravitating, infinite, homogeneous system  was first studied by Jeans (1929) in the context of a gaseous nebula. The resulting Jeans criterion for the onset of gravitational instabilities is applied commonly in a variety of astrophysical systems.
Most astrophysical systems, however, are not truly isolated but are under the influence of an external field, such as the tidal field due to a neighbour. In this case it is not correct to use the standard Jeans criterion to decide whether or not the system is unstable to the growth of instabilities. The presence of an external tidal field is likely to play a major role in the formation of instabilities, but
surprisingly, this idea has not received the attention it deserves.
The tidal stability of an existing bound feature is routinely examined in astronomy, but the effect of a tidal field for the onset or on a growing instability is not considered. 
Obviously this would play a more important role on a growing instability than on a feature that is already formed. 

Section 2 contains the derivation of the modified dispersion relation and the results, the 
applications  are given in Section 3, and  Sections 4 and 5 contain discussion  and the conclusions respectively. 

\section {Formulation of equations}

\subsection  {Linear perturbation analysis: Jeans Criterion}
First we summarise the standard Jeans analysis 
to study the stability of a homogeneous, self-gravitating medium supported by pressure (e.g., Rohlfs 1977). Let $\rho$, denote the density and  $c$ the one-dimensional r.m.s velocity dispersion or the sound speed of this fluid.
 The linearised perturbation equations for the continuity equation, the force equation, and the Poisson equation are given respectively by
$$  \partial {\rho_1} /  \partial t \: \: + \: \rho_0 \: {\nabla} . \: v_1  = 0 \: , \eqno(1) $$
$$ \rho_0 \: \: \partial v_1 / \partial t = - \rho_0 \: \:  \nabla \: \Phi_1 - \nabla (c^2 {\rho_1}) \: ,  \eqno (2)$$
$$ {\nabla}^2 \: {\Phi}_1 \: = \: 4 \pi G \rho_1 .  \eqno (3) $$
Here $\rho_1$, $\Phi_1$ and $v_1$ denote the perturbed density, potential and the radial velocity respectively. These equations can be combined
by taking the divergence of equation (2) and the time derivative of equation (1) to give
$$ {\partial}^2 \rho_1 / \partial t^2 \: - \: c^2 {\nabla}^2 \rho_1 - 4 \pi G \rho_0 \rho_1 = 0 \:  . \eqno (4) $$
This is a wave equation and can be solved by the method of modes.
The perturbed density, $\rho_1$ is 
taken to have   a form like a wave and is proportional to exp $ i (wt - kr)$
where $w$ is the frequency of the perturbation and $k$ is its wavenumber.
Thus equation 4 reduces to
 the familiar dispersion relation
$$ w^2 = k^2 c^2 - 4 \pi G \rho_0 \: , \eqno (5) $$
The system is unstable to the growth of linear perturbations if $w^2 < 0$. This occurs
 for a scale greater than the Jeans length ${\lambda}_c$, where
$${\lambda}_c =  ( \pi c^2 / G \rho_0 )^{1/2} \: . \eqno (6) $$
The corresponding mass within a sphere of radius equal to half the Jeans length $(1/2){\lambda}_c$ is defined to be the Jeans mass, $M_J$ (Binney \& Tremaine 1987) as
$$ M_J = (4 \pi/3) \rho_0 ({\lambda}_c /2)^3  =   (\pi \rho_0 /6)  (\pi c^2 / G \rho_0)^{3/2} \: .  \eqno (7) $$
\noindent In deriving the linearised force equation, it has been assumed that $(\rho_0 + \rho_1) 
\: \nabla \: (\Phi_0 + \Phi_1) $ (in the r.h.s. of equation 2) can be set equal to $ \rho_0 \: \nabla \Phi_1$, that is, $\nabla \Phi_0 = 0$.
This is the so-called Jeans swindle which assumes that the unperturbed potential is constant over the region of interest (which is taken to be local) and hence its derivatives can be ignored (e.g., Binney \& Tremaine 1987). 
The work in this paper does not involve  the usual assumption as in the Jeans swindle, since the  gravitational field is taken to vary over the region of interest and is imposed  externally which itself does not get disturbed. This is different from the growth of perturbation of the medium itself studied in the Jeans analysis.

\subsection {Modification due to an external field}

Next consider the case when the above medium is under an external field with potential given by
$\Phi_{ext}$. The equations derived below including the dispersion relation are general, but we concentrate on the case when the external field is of gravitational origin.
The presence of the external field adds an extra term in the force equation, namely, 
($\rho_0 + \rho_1) \nabla \Phi_{ext}$. Thus, the linearized perturbed force equation  then  becomes
$$ \rho_0 \: \: \partial v_1 / \partial t = - \rho_0 \: \:  \nabla \: \Phi_1 - \: \nabla (c^2 {\rho_1}) - \: \rho_1 \: \nabla \Phi_{ext} \: . \eqno (8)$$
Note that this is valid for a general case and does not assume the external potential to be constant (unlike the 
assumption of constant potential inherent in the Jeans approach), nor does it assume the external field to be a minor perturbation.

 The continuity equation and the Poisson equation for the perturbed motion (equations  1 and 3 respectively) are not affected by the imposed field which only acts externally.
Combining equations 1, 3 and 8, and using the method of modes with  $\rho_1$ taken to be proportional to 
$exp \: \:  i(wt - kr)$, gives the following modified dispersion relation
$$ w^2 =  k^2 c^2 - 4 \pi G \rho_0  -  \frac {{\partial}^2 \Phi_{ext}}{\partial r^2} =  k^2 c^2 - 4 \pi G \rho_0  
+ T_0 \: , \eqno (9) $$
\noindent where 
$T_0$ is the external tidal force per unit distance along the radial direction
$$T_0 = - {{\partial}^2 \Phi_{ext}}/{\partial r^2} \: .  \eqno(10)$$
Thus, the external field results in an additional term in the dispersion relation whose physical origin is transparent. 
 Even though a tidal field is anisotropic, the above dispersion relation is  a  good approximation for the 3-D case, as shown in Section 4. We note this dispersion relation is also valid for other external fields.

\subsection {Results in an external field}
Some noteworthy features of this modified dispersion relation and its solution are as follows:

\noindent  1. In the limit when the tidal field magnitude is zero, 
the modified dispersion relation reduces to the standard case, as expected physically.

\noindent 2.  The tidal field is disruptive, when  $ T_0 >  0$. Here the tidal field opposes the self-gravity of the perturbation. A simple example of the disruptive field is due to a point mass M$_p$, which gives T$_0$ = 2 G M$_p / r^3$ ($>0$). 
In this case, even when the system by itself is unstable to the growth of perturbations,
the addition of the tidal term in equation 9 tends to make it more stable.
Thus, the modified Jeans length, $({\lambda}_c)'$ above which the system becomes unstable 
is larger than the standard Jeans length, 
$\lambda_c$ (equation 6),  and 
is given by
$$({\lambda}_c)'  
=  \lambda_c \: \frac{1}{(1 - T_0 / 4 \pi G \rho_0)^{1/2}}  \: .  \eqno (11) $$
In this case, the minimum mass that can support instabilities is higher (super-Jeans)
  than the standard Jeans mass, $M_J$ (see equation 7), and is given by
 $$  (M_J)_{super} = (4 \pi/3) \rho_0 {{({\lambda}_c}' / 2)}^3 =  {\frac {M_J} {(1 - T_0 /4 \pi G \rho_0)^{3/2}}}  \: .
   \eqno (12) $$
\noindent 3. Conversely, under some conditions the tidal field could be compressive ($T_0 < 0$), which would tend to make the system unstable even when it is stable by itself. A compressive tidal field may seem counter-intuitive but can typically occur when the external field is smooth and varies slowly with radius on scales of the perturbation
 and thus enhances the 
self-gravity of the perturbation (e.g.,  Ostriker, Spitzer \& Chevalier 1972, Das \& Jog 1999, Masi 2007, Renaud et al. 2009).
Define the effective density $\rho_{eff}$ in terms of the compressive tidal field as
$$ \rho_{eff} = \vert T_0 \vert / 4 \pi G  \: . \eqno(13) $$
\noindent This adds on to the gas density $\rho_0$ in equation 9, thus a compressive field makes the system prone to being unstable.

For a compressive field, the resulting critical wavelength that can support gravitational instabilities is lower, 
given by equation (11) with T$_0 < 0$. The  
 minimum mass that can support instabilities is then lower than usual (sub-Jeans)
$$  (M_J)_{sub} = (4 \pi/3) \rho_0 {{({\lambda}_c}' / 2)}^3 =  {\frac {M_J} {(1 + \vert T_0 \vert /4 \pi G \rho_0)^{3/2}}}
  \: .   \eqno (14) $$
\noindent 4. The growth time for perturbations much larger than the Jeans length is 
modified from the usual $1/(4 \pi G \rho_0)^{1/2}$ (see equation 5) by a factor
 $1/(1-T_0/4 \pi G \rho_0)^{1/2}$ (equation 9).  Thus the perturbations grow slower in a disruptive field, and faster in a  compressive field, compared to the normal case.
\section {Applications}
Since the external tidal fields are ubiquitous,  it is important to study their effect.
Here we discuss some astrophysical systems as illustrative examples where the tidal field is likely to affect the formation of gravitational instabilities.
\subsection {Disruptive tidal field}
A typical example is a cloud in a galaxy which experiences the disruptive tidal field of the 
galactic disc and the dark matter halo. 
Recall that for a uniform density
cloud to be tidally stable in the external field, the condition 
$ (4/3) \pi G \rho_0 > T_0$ has to be satisfied, where $T_0 = 4 A (A-B)$ is the tidal field across the cloud, given
in terms of the local Oort $A$ and $B$ constants (e.g., Mihalas \& Routly 1968). Using the  values of $A = 14.5 $ km s$^{-1}$ kpc$^{-1}$ and $B = -12 $ km s$^{-1}$ kpc$^{-1}$ (Binney \& Tremaine 1987), the ratio of the self-gravitational term to the tidal term in the above expression is 1.3 $\times \rho_0 $ (in  H$_2$ cm$^{-3}$). This is $>>$1 since the average
density within a molecular cloud 
 $\geq$ 100 H$_2$ cm$^{-3}$ (Scoville \& Sanders 1987). 
Thus a molecular cloud is highly stable against the galactic tidal field. Note that this is also the condition  that the tidal field does not affect the
growth of instabilities in the molecular cloud (see equation 9).
Conversely, when the tidal field dominates over the self-gravity, not only are such objects tidally unstable but these cannot form at all as gravitational instabilities.
It should be noted that a real molecular cloud is also affected by other physical processes such as turbulent support, thermal pressure (McKee \& Ostriker 2007, Hennebelle \& Falgarone 2012) which would affect its net stability.

When the self-gravity dominates over the tidal field, the instabilities can grow but at a smaller rate than in the absence of the tidal field, and the corresponding Jeans wavelength would be higher.
Thus an external tidal field, even when it cannot fully prevent instabilities from occurring, will still hinder
these as reflected in their lower growth rate.
However, if the magnitude of $T_0$ is small, it cannot affect the stability of the region significantly.

Thus although the resulting dispersion relation (equation 9) is general, its applicability is limited to a certain range of the
tidal field. The tidal field effect will be a maximum when the magnitude
of the disruptive tidal field is close to (but smaller than) the self-gravitational force term.
Thus, the galactic field could play a role in affecting the stability of  low density clouds observed in the outskirts of a galaxy.

Another application could be to the internal structure of clouds. The clouds typically contain denser clumps (e.g., Scoville \& Sanders 1987, Myers 2011) but since these presumably
 evolved as non-liner perturbations from the parent cloud, the "external" field of the parent cloud cannot have much effect on the denser clumps.
However, a clump could be affected by the cumulative tidal field of many nearby clumps, and this could prevent growth
of instabilities or star formation in the clumps. This could
 explain the super-Jeans but starless cores that are observed (Sadavoy et al. 2010)
in a crowded field in a dense cloud as in Orion.

\subsection {Compressive tidal field}
When the external field is compressive 
the system is prone to being unstable (Section 2.3). 
Some astrophysical applications of a compressive tidal field are as follows:

\medskip

\noindent {\bf Centres of early-type galaxies:}
The mass distribution in centres of  elliptical or early-type S0 galaxies
can be described by a spherical potential, as in the 
one-parameter $\eta$ model (Tremaine et al. 1994).
The corresponding tidal field 
was calculated by Das \& Jog (1999) to be
$$  T_0  = -{\frac {G M_g}{{r_0}^3}} {\frac
{{r'}^{(\eta - 3)}}{(1+r')^{\eta + 1}}} (\eta - 2 - 2 r') \: , \eqno (15)$$
\noindent where $r' = r / r_0$, $r_0$ is the core or observed break radius,  $M_g$ is the mass of the galaxy, and $ 0 < \eta \leq 3$.
For $0 < \eta \leq 2$ the mass distribution
at the centre is cuspy,
 and the tidal field is disruptive at all radii. In contrast,
 for the range $2  < \eta \leq 3$ the mass distribution shows a central flat-core distribution, and the tidal field is compressive in the inner parts at radii $ r < r_0 / 2$
for $\eta = 3$ used here. 

For a sample of early-type, flat-core galaxies studied using HST by Faber et al. (1997), the values of the break radius and the mass of the galaxy were obtained (Das \& Jog 1999). Using these, we next obtain the values of the tidal field, and the associated effective density (equation 12) which adds to the true gas density, at $r = 1.4 r_0$.
\begin{table}
\centering
  \begin{minipage}{140mm}
   \caption{Compressive tidal field: Early-type, flat-cored galaxies} 
\begin{tabular}{llll}
Galaxy Name& r$_0$,core radius & Galaxy mass& $\rho_{eff}$ \footnote{Effective density due to compressive field, $\rho_{eff}=\vert T_0 \vert / 4 \pi G$}
 \\
& (in pc) & (in 10$^{11}$ M$_{\odot}) $  & (in H$_2$cm$^{-3}$) \\
&&&\\
NGC 4472 &  178 & 8.4& 9.8$\times 10^3$ \\
NGC 4649& 263& 9.9 & 3.6$\times 10^3$ \\
NGC 1400 & 35 & 2.4 & 3.7$\times 10^5$ \\
NGC 1316 & 36& 2.9 &  4.1$\times 10^5$ \\
NGC 1600 & 759 & 15 &  0.2$\times 10^3$ \\
NGC 3379 & 83 & 1.0 &  1.1$\times 10^4$ \\
NGC 4486 & 562 & 9.0 & 0.3$\times 10^3$ \\
\end{tabular}
\end{minipage}
\end{table}
Table 1 shows that the typical values of the effective central density 
are $\sim 10^3 -10^5$ H$_2$ cm$^{-3}$. Recently a large set of 260 early type galaxies has been studied in the ATLAS$^{3D}$ project (Cappellari et al. 2011) - which also contain four of our sample galaxies- namely NGC 4472, NGC 4649, NGC 3379, and NGC 4486. Eighteen galaxies of the above sample are measured to have typical central molecular gas densities of $10^3 - 10^4 $ H$_2$ cm$^{-3}$  (Bayet et al. 2013). These are much smaller than $\rho_{eff}$. Thus the
compressive field in the centres of early-type flat-core galaxies 
 strengthens and in most cases even dominates the gas self-gravity, and thus can 
favour star formation with sub-Jeans masses. 

This can explain in a natural way why the luminosity in the central regions of flat-core galaxies is observed to be high (Faber et al. 1997). We predict that nuclear
star clusters can form in these galaxies, as is observed (e.g., Boker et al. 2002, Boker 2010). The mass of instabilities would depend on the gas properties and the tidal field strength, and needs to be checked by future detailed work.  We caution that true star formation is not set solely by the onset of instabilities as discussed so far. Rather, it is also affected by other processes such as turbulence, outflows etc (MacLow \& Klessen 2004, McKee \& Ostriker 2007).

Our work gives further justification for the idea proposed by Emsellem 
\& van de Ven (2008) that compressive tidal fields in flat-core galaxies with a small value of the Sersic index can give rise to nuclear star clusters. 
We stress, however, that the actual magnitude of the tidal field is crucial, so that if it is much smaller then the self-gravity of the perturbation ($\rho_{eff} << \rho_0$) it cannot be of much importance in  the formation of instabilities. Thus the field being compressive is not sufficient to result in nuclear star clusters as assumed by Emsellem \& van de Ven (2008).

Further, if the central star cluster evolves into a black hole
 or is accompanied by a nuclear black hole, its effect acts in an opposite way since
the net tidal field may no longer be compressive. This confirms the results on the stability of a circumstellar disc around a central black hole which shows a disruptive field (Kawata, Cen \& Ho 2007). Thus, the compressive field cannot give rise to the central massive object or a black hole.

\medskip

\noindent {\bf  Region of rising rotation curve:} The central region of galaxies with rising rotation curves, including the
special case of constant density or solid body rotation region, 
shows a compressive tidal field as shown by Das \& Jog (1999). Consider a simple, idealized case where the rotation 
velocity $V(R)$  rises to a value $V_0$ at a galactocentric radius $R_d$ and is then flat beyond that radius 
$$ V(R) = V_0 \: (R/R_d)^{\beta}  \eqno (16) $$
\noindent where $R$ is the cylindrical co-ordinate,  $\beta$ is a  power law index $(\neq 0)$ for $R < R_d$,  and $\beta =0$ for $R \geq R_d$. 
The corresponding potential, and the tidal field along the radial direction, 
for $R < R_d$ were calculated (Das \& Jog 1999) to be
$$\Phi_{ext} = \frac{V_0^2}{2\beta} \left (\frac {R}{R_d} \right)^{2\beta} \: , \eqno(17) $$
and,
$$ T_0 = - (2 \beta - 1) \frac {V_0^2}{R_d^2} \left (\frac {R}{R_d} \right)^{2\beta - 2} \: . \eqno(18) $$
Thus, the tidal field  is compressive for $\beta > 1/2 $. This includes the specific  case of
a linearly rising rotation curve where $\beta =1$ that was assumed by Downes \& Solomon (1998) to model the centres of ultraluminous galaxies.

 The effective density due to this field (using equation 13)  is calculated 
for a sample of ultraluminous galaxies from Downes \& Solomon (1998), see Table 2.
\begin{table}
\centering
  \begin{minipage}{140mm}
   \caption{Compressive tidal field: Rising rotation curves} 
\begin{tabular}{lllll}
Ultraluminous galaxy & R$_d$ & V$_0$ & $\rho_{eff}$\footnote{Effective density due to compressive field, $\rho_{eff}=\vert T_0 \vert / 4 \pi G$}
 & $\rho_0$\footnote{$\rho_0$, observed density (Downes \& Solomon 1998)}\\

& pc& km s$^{-1}$ & H$_2$cm$^{-3}$&  H$_2$cm$^{-3}$\\
02483+4302& 270 & 270& 285&250 \\
VII ZW 31& 290 & 290 & 285&450 \\
Mrk 231 & 75 & 345 & 6000&3600 \\
Arp 193&220&230&310&550 \\
Mrk 273&70&280&4535&1800 \\
Arp 220&200&330&770&900 \\
23365+3604&340&260&165&200 \\
\end{tabular}
\end{minipage}
\end{table}
Again $\rho_{eff}$ is comparable to or higher by a factor of 2-3 than the observed gas
density in the region (last column, Table 2) - though this ratio is somewhat smaller than the case of early-type galaxies. Their effect is additive (see equation 9).
Thus, the compressive field  can favour star formation in the molecular gas in the central region. This would be an
additional triggering mechanism different from the one involving shock-compression of 
gas as proposed by Jog \& Solomon (1992) and Jog \& Das (1992).  Note that the
actual triggering of starbursts is more complex where other physical processes such as turbulence, outflows, and feedback could be important (e.g., Ostriker \& Shetty 2011, Renaud, Kraljic \& Bournaud 2012).

In general, the central regions of a normal galaxy will also show compressive tidal field but of a much smaller  magnitude. For a typical V$_0$ = 250 km s$^{-1}$ and a turnover radius of a few kpc as in our Galaxy, the 
tidal field magnitude is smaller by a factor of $\sim$ 100, so it  cannot
 trigger central starbursts.
Interestingly, the tidal field is compressive over the entire disc for a dwarf, late-type galaxy which typically shows a rising rotation curve till the outermost regions
studied. This  would have implications for early evolution of galaxies which should be explored further.

The idea of compressive tides has been proposed to explain the starbursts seen in
N-body simulations of interacting galaxies (Renaud et al. 2009), however the detailed physics of how the compressive tides actually affect the star formation was not discussed. 
This underlying physics is
provided by the modified dispersion relation and the results for the effective Jeans mass 
as obtained in the present paper. Interestingly, Renaud et al. (2008) showed a correlation between the location of the young star clusters and that of the compressive tides in mergers of galaxies. 
Although this is not a conclusive proof of the role played by compressive tides, it strongly supports the result from the present paper.

\section {Discussion}

\noindent  The Jeans analysis and its extension in this paper are based on a 1-D formalism, even though the Jeans analysis is usually applied to 3-D systems which are isotropic. In general, the tidal field is not isotropic (e.g., Das \& Jog 1999). 
We argue  below that despite this, the formalism and the results from this paper are valid.

In order for the field to be considered compressive, T$_0$ must be $< 0$ along all the three axes. Note that the tidal fields  along the two directions orthogonal to the radial direction
are always $< 0$ (e.g., Das \& Jog 
1999) but their magnitude is different from the radial case.
The tidal field magnitude per unit distance along the normal to the radial 
direction in a spherical potential is given by $[- (1/r) (d \Phi_{ext} / dr)]$, while along the radial direction it is given by ($ - d^2 \Phi_{ext}/ dr^2$) - see equations 2-4  in Das \& Jog (1999). In  
the case of a linearly rising rotation curve (Section 3.2) the two are identically equal, 
while in the case of point mass they differ by a factor of two, and in other cases they differ by a factor of few. 

     Strictly speaking, the onset of instabilities as given by the 
dispersion relation (equation 9)  is applicable for a 3-D system only for an isotropic tidal 
field. However, 
since the tidal fields in the other directions are always compressive and 
have magnitudes comparable to the radial case, this dispersion relation is a good approximation for the 3-D case.  Second,  we can be justified in neglecting this 
anisotropy while applying to gas because of its collisional nature,  so that the  effect 
of the three tidal field components would be felt in an average sense. Thus, the modified dispersion 
relation can be taken to be valid, though with the above caveats.

     For the same reasons, the resulting instabilities and their application to 
astrophysical objects can be taken to be a 
reasonably good approximation. The true instabilities will differ only by a 
factor of few compared to that given by the modified dispersion relation (equation 9) that was
obtained using the radial tidal field.
  
    Thus the simple, easy-to-use  modified Jeans criterion (Section 2) can be 
applied to study  the onset of gravitational instabilities in presence of anisotropic tidal fields.

\section {Conclusions}

We study the linear stability of a homogeneous
gravitating system under the influence of an external gravitational tidal field, and obtain a generalized dispersion relation that governs the growth of linear perturbations in this case. We show that a disruptive tidal field increase the minimum mass that can collapse while in contrast a compressive tidal field allows instabilities to grow even at sub-Jeans length and mass. 

Generally tidal fields are likely to be disruptive in the mid to outer parts of most gravitating systems, while they are likely to be compressive in the central region of flat-core or near-constant density distribution.
As a result of these tidal fields, we are more likely to see larger instabilities with super-Jeans length and mass in the outer parts of galaxies, while in the inner parts one would see instabilities more readily and covering lower mass values.
This can explain in a generic way that star clusters are likely to form in centers of early-type, flat core galaxies,
and how starbursts are triggered in centers of ultraluminous galaxies.

\medskip

\noindent {\bf  Acknowledgements:}  

I  would like to thank the referee, Florent Renaud, for  insightful and constructive comments on the paper, and in particular for raising the 
question about the effect of anisotropy of the tidal field.

\medskip

\noindent {\bf References}

\medskip

\noindent Bayet, E. et al. 2013, MNRAS, 432, 1742

\noindent Binney, J.,  Tremaine, S. 1987, Galactic Dynamics, Princeton Univ. Press, Princeton

\noindent Boker,T., Laine, S., van der Marel, R.P., Sarzi, M., Rix, H.-W., Ho, L.C., Shields, J.C. 2002, AJ, 123, 1389

\noindent Boker, T. 2010, in "Star clusters: Galactic building blocks through time and space", IAU Symposium 266, (eds. R. de Grijs and J. Lepine). Cambridge Univ. Press, Cambridge, p. 58

\noindent Cappellari, M. et al. 2011, MNRAS, 413, 813

\noindent Das, M.,  Jog, C.J. 1999, ApJ, 527, 600

\noindent Downes, D.,  Solomon, P.M. 1998, ApJ, 507, 615

\noindent Emsellem, E., van de Ven, G. 2008, ApJ, 674, 653

\noindent Hennebelle, P., Falgarone, E. 2012, A\&A Rev, 20, 55

\noindent Faber, S.M.  et al. 1997, AJ, 114, 1771

\noindent Jeans,J. 1929, Astronomy \& Cosmogony, 2nd Edition. Cambridge Univ. Press, Cambridge

\noindent Jog, C.J.,  Solomon, P.M. 1992, ApJ, 387, 152

\noindent Jog, C.J., Das, M. 1992, ApJ, 400, 476

\noindent Kawata, D., Cen, R.,  Ho, L.C.  2007, ApJ, 669, 232

\noindent Masi, M. 2007, Am. J. Phys., 75 (2), 116

\noindent  MacLow, M.-M., Klessen, R.S. 2004, Rev.Mod.Phy., 76, 125

\noindent McKee, C.F., Ostriker, E.C. 2007, ARAA, 45, 565

\noindent Mihalas, D., Routly, P.M. 1968, Galactic Astronomy. Freeman, San Fransisco

\noindent Myers, P. 2011, ApJ, 743, 98

\noindent Ostriker, E.C., Shetty, R. 2011, ApJ, 731, 41

\noindent  Ostriker, J.P., Spitzer, L., Chevalier, R.A. 1972, ApJ, 176, L51

\noindent  Renaud, F., Kraljic, K., Bournaud, F. 2012, ApJ, 720, L16

\noindent Renaud, F., Boily, C. M., Naab, T. \&  Theis, Ch. 2009, ApJ, 706, 67

\noindent  Renaud, F., Boily, C.M., Fleck, J.-J., Naab, T., Theis, Ch. 2008, MNRAS, 391, L98

\noindent Rohlfs, K. 1977 Lectures on Density Wave Theory. Springer-Verlag, Berlin

\noindent Sadavoy, S.I., Di Francesco, J., Johnstone,D. 2010, ApJ, 718, L32-L37 

\noindent Scoville, N.Z., Sanders, D.B. 1987, in Interstellar Processes, eds. D.J. Hollenbach and H.A. Thronson. Reidel, Dordrecht, p. 21

\noindent Tremaine, S.D. et al. 1994, AJ, 107, 634

\end{document}